\documentclass[12pt]{article} 

\usepackage{makeidx}
\usepackage{epsfig}
\usepackage{subfig}
\usepackage{graphics}
\usepackage{colortbl}
\usepackage{colordvi}
\usepackage{verbatim}
\usepackage{amsmath, amsthm}
\usepackage{hyperref}
\usepackage{enumerate}
\usepackage{wrapfig}
\usepackage{dcolumn}
\usepackage{bm}
\usepackage{bbm}
\usepackage{amssymb,mathrsfs}

\begin{document}

\title{Path Integral representation for Polymer Quantized Scalar Fields}

\author{Nirmalya Kajuri{\footnote{nirmalya@imsc.res.in}}\\ 
\small Institute of Mathematical Sciences\\  
\small CIT Campus, Taramani, Chennai-600113\\
\small India }

\date{ }
\maketitle
\begin{abstract} According to loop quantum gravity, matter fields must be quantized in a background independent manner. For scalar fields, such a background independent quantization is called polymer quantization and is inequivalent to the standard Schrodinger quantization. It is therefore important to obtain predictions from the polymer quantized scalar field theory and compare with the standard results. As a step towards this, we develop a path integral representation for the polymer quantized scalar field. We notice several crucial differences from the path integral for the schrodinger quantized scalar field. One important difference is the appearance of an extra summation at each point in the path integral for the polymer quantized theory. A second crucial difference is the loss of manifest Lorentz symmetry for a polymer quantized theory on  Minkowski Space. 
\end{abstract}

\section{Introduction}
 In describing most situations in nature, quantum field theories defined on a given background spacetime are adequate. When a background spacetime is itself not well-defined, as in the early universe or in the last stages of the evaporation of a black hole, such a description will not do and a more fundamental description seems necessary. According to Loop Quantum Gravity \cite{rovelli, ashtekar, thiemann}, to obtain an appropriate description of such situations we must quantize both gravity and matter fields in a background-independent manner. For scalar fields such a background-independent quantization was develped by Ashtekar et al\cite{polymerfock} and is known as polymer quantization. The theory one obtains is necessarily inequivalent to the standard Schrodinger quantized theory obtained via background-dependent quantization. As we know that the Schrodinger quantized theory is adequate for fixed backgrounds it is important to understand the relationship between the two theories.

In this work, we take a step towards this by obtaining a path integral representation for a polymer quantized scalar field theory on Minkowski spacetime. Let us clarify what we mean by 'on a Minkowski spacetime' here as we have just claimed that the quantization is background independent. Indeed, the construction of the polymer Hilbert Space is a background independent procedure and does not require any   information about the geometry of the background spacetime. It is only through the Hamiltonian that such information enters. Thus by 'on Minkowski spacetime' we mean that we choose the form of the operator to be that for a scalar field theory in a Minkowski background. In particular we will consider the free Klein Gordon Hamiltonian in this paper, although the derivation of the path integral presented here goes through for any arbitrary polynomial term (in the field or its derivatives) term added to this Hamiltonian. To summarize, we will consider the Hilbert Space to be polymer quantized and the Hamiltonian to be the free Klein Gordon Hamiltonian on Minkowski Space. 

 For this derivation, we'll follow the strategy adopted in usual field theory treatments (see for instance,\cite{greiner}). We'll start from a polymer quantized scalar field theory defined on a regular lattice, obtain a path integral representation for the transition amplitude on this lattice and finally take a continuum limit.

One distinctive feature of the polymer quantized scalar field theory is that both field and its conjugate momentum cannot be represented as operators on the polymer Hilbert Space. We may choose one or the other to be represented as well-defined operators. Here we will choose the conjugate momentum to be well-defined. We will see that we may define an \textit{approximate} field operator through the introduction of a scale $\mu$. The definition of the Hamiltonian must then also involve this approximate field operator. The consequence of this in the path integral representation is a modification of the action which, as we shall see, results in the loss of manifest Lorentz symmetry (For a different approach to polymer quantization of scalar field theory which also modifies Lorentz invariance, see \cite{viqar}).  

This scale $\mu$ can be accurately thought of as a lattice spacing in the space in which the conjugate momentum takes values. Defining the approximate field operator with respect to such a scale has the consequence of restricting us to one such lattice. As the momentum takes values on a lattice, its conjugate field takes values on a circle. The polymer quantized scalar field theory is therefore a sigma model with the circle as target space. We will see that the upshot of this in the path integral representation is that the 'path integral' now includes an extra summation at each point .This is another novel feature of the polymer scalar theory that does not occur in the usual Schrodinger quantized scalar field theory.

The paper is organized as follows. In section I we'll demonstrate our strategy by obtaining a path integral representation for a simpler system, the polymer quantized simple harmonic oscillator. Section II   briefly recalls polymer quantized scalar field theory and introduces some new notation. In section III the path integral representation for the polymer quantized scalar field is derived. In section V we conclude with a discussion of the results. 
\section{Path Integral formulation of the polymer quantized SHO}

In this section we'll obtain the path integral representation for the relatively simple case of a  simple harmonic oscillator in the polymer representation of quantum mechanics. The polymer representation of quantum mechanics \cite{polymer1, polymer2, shadow} is a quantization method for the non-relativistic point particle which closely resembles polymer quantization of scalar fields (and loop quantization of gravity).   It had been introduced in \cite{shadow} as a simple toy model to explore certain features of Loop Quantum Gravity. This model will be useful to us to demonstrate our strategy for obtaining a path integral for the polymer quantized scalar field in a simple setting. Moreover, the features that distinguish the path integral representation of polymer scalar field theory from that of the usual scalar field theory will already appear here.

We begin with an introduction of the model.

\subsection{ Simple Harmonic Oscillator in the polymer representation}
  Like the standard Schrodinger representation, the polymer representations of quantum mechanics   are based on  unitary representations of the Weyl Algebra, which is given by: \begin{align}
\nonumber & U({\lambda_1}) U({\lambda_2}) =  U({\lambda_1 + \lambda_2}) 
\\&  \nonumber  V({\mu_1}) V({\mu_2}) =  V({\mu_1 + \mu_2})   
\\ & \nonumber  U({\lambda})V({\mu}) = e^{-i{\lambda}{\mu}}V({\mu}) U({\lambda}) 
\end{align} 

Unlike the Schrodinger representation however, the representation of the Weyl algebra in the polymer representations are not weakly continuous \footnote{ An operator $U(\mu)$ is weakly continuous when all matrix elements of $U(\mu)$ are continuous in $\mu$}. In the Schrodinger representation $U(\lambda), V(\mu)$ can be understood as exponentiated position and momentum observables, respectively. Since both their representations cannot be weakly continuous in polymer quantization, both momentum and position cannot be well-defined operators. There are two possible polymer representations depending on which of the two subgroups $U(\lambda), V(\mu)$ is taken to be continuously represented. Here we'll consider the case where the momentum operator is defined.

Let us recall the polymer representation of quantum mechanics in a bit more detail. To construct the Hilbert Space, we first choose a countable set, $\gamma = \{ {p}_j, {p}_j \epsilon \mathbb{R} \}$ and define a set Cyl$_{\gamma}$ of
linear combinations of the form: Cyl$_{\gamma} := \{\sum_{j} f_j e^{i  {p}_j {x}}, f_j \in \mathbb{C}  \}$ The $ f_j $ are subject to certain regularity conditions \cite{shadow}. Then we define the set of functions of $ x$, Cyl := $\cup_{\gamma}$Cyl$_{\gamma}$. The inner product on this set is chosen to be
\begin{align} 
(e^{i{p_i} {x}}, e^{i {p_j} x}) = \delta_{ {p_i},{p_j}} 
\end{align}
  $\{e^{ip {x}}~/p \in \mathbb{R} \}$  form an uncountable basis of this space and we denote them as the kets $|p\rangle$. The completion of Cyl w.r.t this inner product is our requisite Hilbert Space  $H_{\mathrm{poly}}$: $\overline{\mathrm{Cyl}} =: H_{\mathrm{poly}}$. 

On this Hilbert Space we have the basic operators:
\begin{align} 
\hat{p}|p\rangle = p|p\rangle
\end{align}
and
\begin{align}  \hat{V}(\lambda)|p\rangle = |p -\lambda\hbar\rangle \end{align}

As $ \hat{V}(\lambda)$ is not weakly continuous in $\lambda $ a position operator cannot be defined. We can however define an approximate position operator by choosing some scale $ \mu_0 $:  \begin{align}\hat{x}_{\mu_0}=\frac{\hat{V}(\mu_0)-\hat{V}(-\mu_0)}{2\mu_0 i}\end{align}   Once the approximate position operator is defined with a particular choice $\mu_0$, starting from a given $|p_0\rangle $ and acting on it with$\hat{V}(\mu_0) $ we'll generate a set of basis vectors $ \{|p\rangle  = |p_0 + n\mu_0\hbar \rangle \}  $.  This gives a proper subspace of the Hilbert Space and the action of our observables will leave the subspace invariant. We'll be working in one such subspace.

We consider the simple harmonic oscillator with k=2. That is, we take the classical Hamiltonian to be \begin{align} \frac{\hat{p}^2}{2m} +  x^2 \end{align} In polymer representation the Hamiltonian reads:  \begin{align} \frac{\hat{p}^2}{2m} + \left(\frac{\hat{V}(\mu_0)-\hat{V}(-\mu_0)}{2\mu_0 i}\right)^2 \end{align}
\subsection{ Path Integral formulation}

We start by introducing a new set of basis states which we may call the 'approximate position states'. These are given by: 

\begin{align} 
|x\rangle = \sum_n e^{in\hbar\mu_0 x}|p_0 +n\mu_0\hbar \rangle \qquad x\in [0, 2\pi/\mu_0)
\end{align}

These are eigenvectors of the approximate position operator : 
\begin{align} 
\hat{x}_{\mu_0}|x\rangle = \frac{\sin \mu_0 x}{\mu}|x\rangle
\end{align}

The inner product of the approximate position states is given by 
\begin{align} 
\langle x' |x\rangle =  \delta (x'-x)
\end{align}

These are therefore non-normalizable states and do not belong to the polymer Hilbert Space. In this they are similar to the position basis of standard Quantum Mechanics.  We will now obtain a path integral representation of the amplitude of transition between two states belonging to this basis. The advantage of using this basis for the derivation of the path integral representation (as opposed to the momentum basis as in \cite{campiglia}) is that it makes the connection between the polymer quantum mechanics and a particle on a circle in ordinary quantum mechanics transparent. Indeed, the problem is now equivalent to obtaining a path integral representation for a point particle on a circle. In what follows, we'll be adhering closely to the treatment for a point particle on a circle given in \cite{kleinert}.

Our starting point is the transition amplitude $ \langle x_f|e^{- \frac{i}{\hbar}t\hat{H} }|x_i\rangle $.
We divide t into N pieces $\epsilon$ = t/N. So 
\begin{align*}
e^{- \frac{i}{\hbar}{\hat{H}t}} = \Pi^N_{n=1} e^{- \frac{i}{\hbar}{\hat{H}\epsilon}}
\end{align*} 
Inserting complete basis of the form $ \mathbbm{1} =    \int^\frac{2\pi}{\mu_0}_0 |x \rangle \langle x|$ in between each factor we have

 \begin{align} \langle x_f|e^{- \frac{i}{\hbar}\hat{H}t} |x_i \rangle  =   \int dx_{N-1}...dx_1 \langle x_f| e^{-\frac{i}{\hbar}{\hat{H}\epsilon}} |x_{N-1}\rangle.......\langle x_1| e^{- \frac{i}{\hbar}{\hat{H}\epsilon}}|x_i \rangle \label{6}\end{align}

Taking N very large ($\epsilon <<1 $) and expanding the nth term of the series in $\epsilon $ we have 
\begin{align} \langle x_{n+1}| e^{-\frac{i}{\hbar}{\hat{H}\epsilon}} |x_{n}\rangle = \delta (x_{n+1} - x_n) - \frac{i}{\hbar}\epsilon \langle x_{n+1}|\hat{H}|x_{n}\rangle + \mathcal {O} (\epsilon ^2) \label{6.5}\end{align}

The matrix elements of H are:
\begin{align} 
 \langle x_{n+1}|\hat{H}|x_{n}\rangle = \langle x_{n+1}|\frac{\hat{p}^2}{2m} |x_{n}\rangle  +\langle x_{n+1}|{\hat{x}_{\mu_0}}^2 |x_{n}\rangle  
 \end{align} 
\begin{align}
 =  \langle x_{n+1}|\frac{\hat{p}^2}{2m} |x_{n}\rangle +\frac{\sin^2 \mu_0 x_n}{\mu_0^2}  \langle x_{n+1}|x_{n}\rangle  \label{7}
 \end{align}
Now we insert the completeness relation $ \mathbbm{1} = \sum|p \rangle \langle p|$ (where the sum is over the lattice in $ \{|p\rangle  = |p_0 + n\mu_0\hbar \rangle \} ) $\eqref{7} to obtain:
\begin{align}
\nonumber &\sum_{p_{n+1}}\frac{{p_{n+1}}^2}{2m} \langle x_{n+1}|p_{n+1} \rangle \langle p_{n+1} |x_{n}\rangle + \frac{\sin^2 \mu_0 x_n}{\mu_0^2}\langle x_{n+1}|p_{n+1} \rangle \langle p_{n+1} |x_{n}\rangle \\
&\sum_{p_{n+1}}\frac{\mu_0}{2\pi} e^{ip_{n+1}(x_{n+1}-x_n)}\left(\frac{{p_{n+1}}^2}{2m} + \frac{\sin^2 \mu_0 x_n}{\mu_0^2}\right) \label{8}
\end{align}
From \eqref{8}, \eqref{7} and \eqref{6.5} we obtain up to $\mathcal {O} (\epsilon ^2)$ terms:
\begin{align} 
\nonumber  \langle x_{n+1}| e^{-\frac{i}{\hbar}\hat{H}\epsilon}|x_{n}\rangle 
 =  \sum_{p_{n+1}}e^{ip_{n+1}(x_{n+1}-x_n)}  \left[1- \frac{i\epsilon}{\hbar}\left(\frac{p_{n+1}^2}{2m}+  \frac{\sin^2{({\mu_0}x_{n})}}{{\mu_0}^2}\right) \right] 
\end{align}
\begin{align}
 =  \sum_{p_{n+1}} \frac{\mu_0}{2\pi} e^{ip_{n+1}(x_{n+1}-x_n)-\frac{i\epsilon}{\hbar}(\frac{p_{n+1}^2}{2m} +\frac{\sin^2{({\mu_0}x_{n})}}{{\mu_0}^2})} 
\end{align}
 Substituting this in \eqref{6} we have 
\begin{align}
\langle x_f|e^{- \frac{i}{\hbar}\hat{H}t} |x_i \rangle  =  \left(\frac{ \mu_0}{2\pi}\right)^{N}  \sum\limits_{p_{N},....,p_1}\int dx_{N-1}...dx_1 e^{-\frac{i}{\hbar}S_N}  
\end{align}
where
\begin{align} 
S_N = \epsilon\sum\limits_{n=0}^{N-1} p_{n+1}\frac{(x_{n+1}-x_n)}{\epsilon} +\frac{ p_{n+1}^2}{2m} +\frac{\sin^2{({\mu_0}x_{n})}}{{\mu_0}^2}
\end{align}

We may now take the $N\rightarrow \infty$ limit and obtain an expression for the path integral. However this will be very different from the usual form of the path integral, with  $p$ taking only discrete values and $x$ being bounded. However it is possible to express the path integral in a more familiar form. Note that to obtain \eqref{8} from \eqref{7} we had inserted the complete momentum basis. This is equivalent to using the expansion of the $\delta$ function :
\begin{align}
\delta(x_{n+1}-x_n) = \sum_{p_{n+1}=-\infty}^{\infty} \frac{ \mu_0}{2\pi} e^{ip_{n+1}(x_{n+1}-x_n)} \label{old}
\end{align}

However, using Poisson's formula 
\begin{align}
\sum_{l=-\infty}^{\infty} e^{2\pi i kl} = \sum_{m=-\infty}^{\infty} \delta(k-m)
\end{align}

We may rewrite 
\begin{align}
\langle x_{n+1}|x_{n}\rangle = \sum_{l=-\infty}^{\infty} \delta(x_{n+1}-x_{n} +\frac{2\pi}{\mu} l)
\end{align}
Fourier decomposing the $\delta$ function gives:
\begin{align}
\langle x_{n+1}|x_{n}\rangle = \sum_{l=-\infty}^{\infty} \int_{-\infty}^{\infty}\frac{\mu_0}{2\pi} dp_{n+1}  e^{ip_{n+1}(x_{n+1}-x_n) +\frac{2\pi}{\mu} i p_{n+1}l} \label{new}
\end{align}

If we use \eqref{new} instead of \eqref{old} in deriving the path integral expression, the transition amplitude takes the following form: 
\begin{align}
\prod_{n=1}^{N-1} \int_0^{2/pi} dx_n \prod_{n=1}^{N}\left( \frac{\mu_0}{2\pi} \int_{-\infty}^{\infty} dp_n \sum_{l_n=-\infty}^{\infty} \right) e^{i\sum_{n=1}^N p_n(x_n - x_{n-1} + \frac{2\pi}{\mu} l \delta_{n, N}) -\epsilon H(x_n, p_n)}
\end{align}

where $$ H(x_n,p_n) = \frac{ p_{n}^2}{2m} +\frac{\sin^2{({\mu_0}x_{n-1})}}{{\mu_0}^2} $$

Now we absorb each sum over $l_n$ into the corresponding $x_n$ by extending the range of integration from $[0,2\pi/\mu_0)$ to $(-\infty, \infty)$. Note however that the number of sums over $l_n$ is one more than the number of $x_n$ integrations, which means that the last sum $\sum_{l_N}$ remains as it is. We then have the following expression for the transition amplitude: 
\begin{align}
K(x_f,x_i) = \sum_{l=-\infty}^{\infty} \prod_{n=1}^{N-1} \int_{-\infty}^{\infty}  dx_n \prod_{n=1}^{N} \frac{\mu_0}{2\pi} \int_{-\infty}^{\infty} dp_n e^{i\sum_{n=1}^N p_n(x_n - x_{n-1} + \frac{2\pi}{\mu} l \delta_{n, N}) -\epsilon H(x_n, p_n)}
\end{align}

We notice that this expression can be re-written as 
\begin{align}
K(x_f,x_i) = \sum_{l=-\infty}^{\infty} K(x_f + \frac{2\pi}{\mu}l,x_i)_{\text{noncyclic}}
\end{align}
where $K(x_f, x_i)_{\text{noncyclic}}$ is the ordinary path integral with the Hamiltonian $ H(x_n, p_n)$, i.e it does not involve a sum over repated final points. 

Now we take the $N\rightarrow \infty$ limit. This gives:
\begin{align}
K(x_f, x_i)_{\text{noncyclic}} =  \int  [\mathcal{D}p][\mathcal{D}x] e^{\frac{i}{\hbar}S}
\end{align}
 where
\begin{align}
S =   \int_0^t \mathrm{d}\tau  p\dot{x}- \left(\frac{p^2}{2m} +\frac{\sin^2{({\mu_0}x )}}{{\mu_0}^2}\right)
\end{align}
and 
\begin{align}
  \int [\mathcal{D}p] [\mathcal{D}x] = \prod_{n=1}^{N-1} \int_{-\infty}^{\infty}  dx_n \prod_{n=1}^{N} \frac{\mu_0}{2\pi} \int_{-\infty}^{\infty} dp_n 
\end{align}
As the momentum integral is Gaussian one may integrate it out\footnote{ That the momentum integral is gaussian is a consequence of our having chosen the polymer representation in which the momentum is well defined. Indeed, we had made that choice to ensure that we can express the transition amplitude as an integral over configuration space. In the alternate polymer representation, it would only be the position path integral which would be Gaussian. }. 
The final expression for the transition amplitude is therefore  

\begin{align}
K(x_f, x_i) = \sum_{l=-\infty}^{\infty} K(x_f + \frac{2\pi}{\mu}l,x_i)_{\text{noncyclic}} = \sum_{l=-\infty}^{\infty}\int_{x_i}^{x_f+ \frac{2\pi}{\mu} l} [\mathcal{D}x] e^{\frac{i}{\hbar} \int_0^T \mathrm{d}\tau \frac{m\dot{x}^2}{2} -\frac{\sin^2{({\mu_0}x)}}{{\mu_0}^2}} \label{ampl}
\end{align}

To summarize, we have obtained a path integral representation for the polymer quantized simple harmonic oscillator. As we have noted earlier, polymer quantization along with a choice of scale $\mu_0$ restricts us to a momentum lattice. This is equivalent to converting the configuration space into a circle.  We saw this explicitly through the introduction of the approximate position basis. We could then use techniques utilized in the derivation of path integral on a circle to derive a path integral representation of the polymer quantized harmonic oscillator. The final expression has two differences from the path integral representation of the Schrodinger quantized harmonic oscillator : 

(i) The $x^2$ term in the action has been replaced by $\frac{\sin^2(\mu_0x)}{{\mu_0}^2}$.

(ii) Now there is an extra sum appearing the path integral expression. This sum is over paths with different end points, the different end points being the points  $x_f+2\pi l, l\in \mathbb{Z}$
 where $x_f$ is the original end point. 

We'll see that this features will re-appear in the path integral representation of the polymer quantized scalar field theory.
 
\section{ The Polymer Quantized Scalar Field}

We now describe the polymer quantized scalar field theory. The key feature of this quantization is background independence. The construction of the polymer Hilbert Space does not require any information about the background geometry. Indeed, it does not even require a constinuous background \cite{polymerfock}. We give the detailed construction below. From now on we'll set $\hbar=1 $. We'll follow the notation of \cite{date}. 

First define a vertex set $V = ( \vec{x}_1, \vec{x}_2, \ldots, \vec{x}_n)$ of
finitely many, distinct points $\in \mathbb{R}^3$. The corresponding vector space Cyl$_V$ is generated by basis vectors:  $$\psi_{V,\vec{\lambda}}(\phi) := e^{i\sum_j \lambda_j \phi(\vec{x}_j)}$$
where $\lambda_j$ are \em non-zero\em \hspace{0.1cm}real numbers. Then define Cyl := $\cup_V$ Cyl$_V$ and on Cyl define the inner product 
\begin{align}
\int d\mu(\phi) \psi^{*}_{V, \lambda}(\phi) \psi_{V,\lambda} (\phi) = \delta_{V,V'}\delta_{\lambda, \lambda'}
\end{align} 

Notice that this inner product is \textit{diffeomorphism invariant}. 
 The Cauchy completion of Cyl w.r.t this inner product gives the Hilbert Space $H_{\mathrm{poly}}$: $\overline{\mathrm{Cyl}} =: H_{\mathrm{poly}}$.
The basic operators here are $ \hat{U} (\lambda,\vec{x})$ and $\hat{\pi}(\vec{x})$. The former acts as:

If $\vec{x}$ is not in $\{\vec{x_j}\}$
\begin{align} 
\hat{U}(\lambda,\vec{x}) e^{ i\sum_j \lambda_j\phi(\vec {x_j})} = e^{ i\sum_j \lambda_j\phi(\vec{x_j})+\lambda\phi(\vec{x})} \label{Udef1}
\end{align} 
 If $ \vec{x} =\vec{x_i} $ $\in$ $\{\vec{x_j}\}$ and $ \lambda_j+\lambda \neq 0 $
\begin{align}\label{Udef2}\hat{U} (\lambda,\vec{}x)e^{ i\sum_j \lambda_j\phi(\vec{x_j})} = e^{ (i\sum_j \lambda_j+\lambda\delta_{\vec{x_j},\vec{x_i}})(\phi(\vec{x_j}))} 
\end{align} 
If $\vec{x} =\vec{x_i} $ $\in$ $\{\vec{x_j}\}$ and $ \lambda_j+\lambda = 0 $
\begin{align}\label{Udef3}\hat{U} (\lambda,\vec{x})e^{ i\sum_j \lambda_j\phi(\vec{x_j})} = e^{ (i\sum_j \lambda_j(\phi(\vec{x_j}))} 
\end{align} 
where the sum is over the set $\{\vec{x_j}\} - \vec{x_i}$.
and the action of the latter is given by
 \begin{align}\hat{\pi}(\vec{x})= \frac{\delta}{\delta \phi(\vec{x})} \end{align}   

Just like the position operator for the polymerised quantum mechanics, the field operator here is not well defined. We may define an approximate field operator using a scale $\mu$:
\begin{align}\hat{ \phi}_{\mu}(\vec{x}) = \frac{ \hat{U} (\mu,\vec{x}) -  \hat{U} (-\mu,\vec{x})} {2\mu i} \label{phimu}
\end{align}
Again, just as in quantum mechanics, the choice of a $\mu$ and a basis vector gives a proper subspace of the Hilbert space on which the basic operators act invariantly. We'll restrict ourselves to one such subspace i.e a lattice in the $\vec{\lambda}$ labels. 

From the presentation given above the connection with polymer quantum mechanics may not be immediately apparent. However this can be remedied by the introduction of a new notation which makes the similarities between the two theories more trasnparent and makes the derivation of the path integral representation straightforward. Let us consider a state $ e^{ i\sum_j \lambda_j\phi(\vec{x_j})}$. We can specify this state by specifying the vertex set V and the values of $\lambda_i$. But we could alternately specify the state by we defining a field $\pi(\vec{x})$ for all $\vec{x}$ such that

$$\pi(\vec{x_j}) = \lambda_j $$ if  $ \vec{x_j}$ $\in $ V 
$$\pi(\vec{x_j}) =  0 $$ otherwise.
In other words the field $\pi(\vec{x})$ has support only on the vertex set. 
Then the same state may be written as $$ e^{ i\sum_{\vec{y}} \pi(\vec{y})\phi(\vec{y})} =: |\{\pi(\vec{y})\} \rangle \leftrightarrow |\pi\rangle $$ 
where the sum is over all values of $y$ . 
 
Now the action of the basic operators maybe represented as   
 \begin{align}
\label{Udef4}
\hat{U} (\lambda,\vec{x})|\{\pi(\vec{y})\} \rangle = |\{\pi(\vec{y})+ \lambda\delta_{\vec{x},\vec{y}}\}\rangle
  \end{align}
and 
\begin{align}
\hat{\pi}(\vec{x})|\{\pi(\vec{y})\} \rangle = \pi(\vec{x})|\{\pi(\vec{y})\}\rangle
 \end{align}
The above equation justifies the choice of labelling these states as $|\{\pi(\vec{y})\}\rangle$ - these are indeed eigenstates of the conjugate momentum operator. Note also that \eqref{Udef4} now incorporates all the cases \eqref{Udef1}, \eqref{Udef2}, \eqref{Udef3}. 

In our notation, we can write the inner product as: 
\begin{align}  \label{delta}
\langle \{\pi(\vec{y})\}|\{\pi'(\vec{y})\}\rangle = \prod_{\vec{y}} \delta_{\pi(\vec{y}),\pi'(\vec{y})}
\end{align}
To understand that this is the same as $\delta_{V,V'}\delta_{\lambda, \lambda'}$ note that this expression equals 1 only if all the values of $ \pi(\vec{x_i})$ and $ \pi'(\vec{x_i})$ agree, that is they should (i) both be non zero on the same points i.e vertex sets V and V' must coincide and (ii) The values of $\lambda$ s must agree on this set. Else it vanishes. 

The advantage of this notation is that the field states can now be thought of as a product of polymer point particle momentum states, with one polymer point particle at each point of the vertex set. To emphasize this, we introduce one final piece of notation. We write $$ |\{\pi(\vec{x})\} \rangle = \underset {\vec{x}}{\prod} |\pi(\vec{x}) \rangle $$ where $ |\pi(\vec{x}) \rangle$ is shorthand for $ e^{i\pi(\vec{x})\phi(\vec{x})}$.

We notice that we may then write \begin{align} 
\mathbbm{1}=   \underset {\vec{x}}{\prod}\underset{\pi'(\vec{x})}{\sum }| \pi'(\vec{x}) \rangle \langle \pi'(\vec{x}) | \label{id}
\end{align}
We verify this using \eqref{delta}:
\begin{align}  \left(\underset {\vec{x}}{\prod}\underset{\pi'(\vec{x})}{\sum }| \pi'(\vec{x}) \rangle \langle \pi'(\vec{x})\right) |\pi\rangle &=  \underset {\vec{x}}{\prod}\underset{\pi'(\vec{x})}{\sum }\delta_{\pi(\vec{x}),\pi'(\vec{x})} e^{i\pi'(\vec{x})\phi(\vec{x})}= \prod_{\vec{x}} e^ {i\pi(\vec{x})\phi(\vec{x})} = |\pi \rangle 
\end{align}

Finally, we noted above that the choice of $\mu$  restricts us to a subspace of the polymer Hilbert Space, just as in the case of polymer quantum mechanics. In the language of our new notation this is equivalent to making the space where the field $\pi(\vec{x})$ takes values (the target space) into a  lattice of spacing $\mu$.

To summarize, we introduced the Hilbert Space for the polymer quantized scalar field theory. The construction was entirely background independent. We saw that as in the case of polymer quantum mechanics, the definition of the approximate field operator required the introduction of a scale $\mu$. We have introduced a new notation which makes the similarity of this theory with polymer quantum mechanics more transparent and which will therefore enable us to follow the strategy of the previous derivation.

\section{Path Integral formulation of the polymer quantized scalar field}

 We'll now obtain a path integral representation for the polymer quantized scalar field theory. The Hamiltonian will be taken to be the usual Klein Gordon Hamiltoninan- $ \int d^3x \frac{1}{2}({\pi}^2(\vec{x}) + (\nabla \phi(\vec{x}))^2 + m^2 \phi^2(\vec{x}) $. However as a field operator is absent we'll have to instead use the approximate field operator in the definition of the Hamiltonian. We lose Lorentz symmetry as a result, as will be made manifest in the path integral representation.

Now we will introduce a basis of approximate field states. These are the analogues of the approximate position states of polymer quantum mechanics. These are defined as: 
\begin{align}
 |\{\phi(\vec{x})\} \rangle = \prod_{\vec{x}} \left( \sum_n e^{in\mu\phi(\vec{x})} |\pi_0(\vec{x})+n\mu \rangle \right), \phi(\vec{x}) \in [0, \frac{2\pi}{\mu}) \forall\vec{x}
\end{align}

Then the inner product between two such states is given by:
\begin{align}
\langle \{\phi(\vec{x})\}  |\{\phi'(\vec{x})\} \rangle = \prod_{\vec{x}} \delta\left( \phi(\vec{x})-\phi'(\vec{x})\right)
\end{align}

And these are eigenstates of $\hat{ \phi}_{\mu}(\vec{x})$, the approximate field operator.
\begin{align}
\hat{ \phi}_{\mu}(\vec{x})|\{\phi(\vec{y})\} \rangle = \frac{ \sin \mu \phi(\vec{x})}{\mu} |\{\phi(\vec{y})\} \rangle 
\end{align}

 Now we'll calculate $ \langle \phi_f| e^ { -i\hat{H}t}|\phi_i \rangle $ by discretizing t into N pieces $ e^{-i\hat{H}t} = \Pi^N_{n=1} e^{-i\hat{H}\epsilon_t} $ Our strategy will be to do this first for a scalar field theory which lives on a lattice and finally take the continuum limit. To this end we discretize space into a cubic lattice with lattice spacing $ \epsilon_x $ and call the entire lattice L from now on. 

So we have
\begin{align}
\langle \phi_f| e^ { -i\hat{H}t}|\phi_i \rangle = \langle \phi_f|\prod^N_{n=1} e^{-i\hat{H}\epsilon_t} |\phi_i \rangle
\end{align}
Using \eqref{id} this is rewritten as:
\begin{align} \langle \phi_f|  e^ { -i\hat{H}t}|\phi_i \rangle =  \prod_{(\vec{x}\in L)}\sum_{\phi_1(\vec{x})}  .....\prod_{(\vec{x}\in L)}\sum_{\phi_{N-1}(\vec{x})} \langle \phi_f(\vec{x})| e^{-i\hat{H}\epsilon_t}| \phi_{N-1}(\vec{x}) \rangle ......\langle \phi_1(\vec{x})|e^{-i\hat{H}\epsilon_t}|\phi_i(\vec{x}) \rangle 
\end{align}

Taking N very large ($\epsilon_t <<1  $) and expanding the nth factor  in $\epsilon_t $ we have 
\begin{align} \langle \phi_{n+1}(\vec{x})| e^{-i\hat{H}\epsilon} |\phi_{n}(\vec{x})\rangle =   \langle \phi_{n+1}(\vec{x}) |\phi_{n}(\vec{x})\rangle  + i\epsilon_t \langle \phi_{n+1}(\vec{x})|\hat{H}|\phi_{n}(\vec{x})\rangle + \mathcal {O} (\epsilon_t ^2) \label{expo} \end{align}

 The classical Klein Gordon Hamiltonian is $ \int d^3x \frac{1}{2}({\pi}^2(\vec{x}) + (\nabla \phi(\vec{x}))^2 + m^2 \phi^2(\vec{x}) $ . We'll use the discretized polymer version, which is given by: 
\begin{align}
H = \sum_{\vec{x}}\epsilon_x^3 \left[ \frac{\pi^2(\vec{x})}{\epsilon_x^2} + {\left( \hat{\alpha} \frac{\phi_{\mu}(x_{\alpha}+\epsilon_x)- \phi_{\mu}(x_{\alpha})}{\epsilon_x^2} \right)}^2 + m^2 \phi_{\mu}^2 (\vec{x})\right]  \qquad \alpha =1,2,3
\end{align}
Here $\hat{\alpha}$ are the unit vectors.
The derivation now proceeds exactly as in the polymer quantum mechanics case and we obtain :
\begin{align}
 \langle \phi_f|  e^ { -i\hat{H}t}|\phi_i \rangle = \prod_{\vec{x} \in L} \left[ \sum_{l=-\infty}^{\infty} \prod_{n=1}^{N-1} \left( \int_{-\infty}^{\infty} d\phi(\vec{x}) \right) \prod_{n=1}^{N} \left(\frac{\mu}{2\pi}\int_{-\infty}^{\infty} d\pi(\vec{x}) \right) \right] e^{i S_{N,L}}
\end{align}

Where 
\begin{align}
S_{N, L} = \sum_L \epsilon_x^3 \sum_n^N \epsilon_t& \left[\left(\frac{\pi_n(\vec{x})(\phi_n(\vec{x})-\phi_{n-1}(\vec{x}) +   \frac{2\pi}{\mu} l \delta_{n,N})}{\epsilon_t} \right)  - \right. \frac{ \pi_n^2(\vec{x})}{\epsilon_x^2} \\ 
&- \left. {\left(\hat{\alpha}\frac {\sin \mu\phi(x_{\alpha}+\epsilon_x) - \sin \mu\phi (x_{\alpha})}{\epsilon_x{\mu}}  \right)}^2 - m^2 \frac{\sin^2 \mu\phi(\vec{x})}{\mu^2} \right]
\end{align}

Taking the continuum limit $\epsilon_t, \epsilon_x \rightarrow 0$ and performing the Gaussian integration over momentum we obtain the final form of the path integral which is 
\begin{align}
K (\phi_f, \phi_i) = \int_{\text{cyclic}}  \mathcal{D}\phi e^{i\int d^3xdt\frac{1}{2}({\dot{\phi}}^2 - (\nabla \phi)^2(cos\mu\phi)^2 -  m^2 (\sin^2{(\mu\phi)}/ {\mu^2}) ) }
\end{align}

Where 
\begin{align}
 \int_{\text{cyclic}}  \mathcal{D}\phi = \prod_{\vec{x}} \sum_{l=-\infty}^{\infty} \int_{\phi_i(\vec{x})}^{\phi_f(\vec{x})+\frac{2\pi}{\mu} l}d\phi(\vec{x})
\end{align}

This is the final form of the  polymer path integral for the scalar field. This is quite similar to the path integral for the polymer quantized harmonic oscillator we derived in section II. In particular the two distinguishing features of the polymer path integral that we noted there appear here as well -

(i) Appearance of the extra summation: There we saw that the position variable took values on a circle. Here it is the field variable which takes values on a circle. The consequence of this in the path integral representation is the presence of an extra summation at each spatial point. 

(ii) Modification of the action:  Here the action is modified from that of the Schrodinger quantized scalar field by the replacement of $x$ by $ \sin{(\mu\phi)}/ \mu $. As a result an extra $\cos^2 (\mu \phi)$ term multiplies the $(\nabla \phi)^2$ present in the action, spoiling the Lorentz symmetry.

\section{Summary and Discussion}

In this work we obtained a path integral representation for a polymer quantized Klein Gordon field. We found that the path integral representation can be put in a form that closely resembles the Schrodinger form except for two differences - (i) Introduction of an extra sum in the path integral and (ii) a replacement of $\phi$ by  $ \sin{(\mu\phi)}/ \mu $ in the action. 

The first modifcation is a consequence of the fact that the field takes values on a circle. We exhibited this fact through our choice of basis \footnote{ See also\cite{Date2}, where it was shown that the classical phase space corresponding to a polymer quantum mechanical theory with one degree of freedom is a cylinder when one restricts to observables defined with a single scale, as we do here.}. This in turn is a consequence of the fact that we defined our observables using a single scale, thereby restricting the values the momentum field may take on a regular lattice. 

The second modification is seen to result in Lorentz violation. The origin of the Lorentz violation is easy to understand. The classical Hamiltonian has terms $(\nabla \phi)^2$ and $\pi^2$, the latter equaling $(\dot{\phi})^2$ in the classical theory. In the transition to the polymer field theory the latter term remains as it is while the latter gets modified as we use an approximate field operator defined with a scale $\mu$ . This spoils the symmetry between the space and time derivatives that existed in the classical theory.

In this work we have succeded in obtaining a path integral for the polymer quantized scalar field theory. We noticed some of the ways in which it differs from the standard path integral for the Schrodinger theory. However to get concrete predictions from the polymer quantized scalar field theory, we need to obtain a perturbative formulation.  Such a formulation will help us ascertain the magnitude of Lorentz violation in the polymer quantized scalar field theory. Our work makes such a study feasible.

\section*{Acknowledgements}
The author is deeply indebted to Ghanashyam Date for many helpful discussions, insights and most importantly for pointing out a crucial error in an earlier draft. The author would also like to thank Alok Laddha for helpful discussions.

\end{document}